\documentclass{appolb}
\usepackage{graphicx}
\begin{document}
\title{Novel Extrapolation for Strong Coupling Expansions
\thanks{Presented at the Strongly Correlated Electron Systems
Conference, Krak\'ow 2002}%
}

% Authors and Affiliations

\author{K.P.\ Schmidt, C.\ Knetter and G.S.\ Uhrig
\address{Institut f\"ur Theoretische Physik,
Z\"ulpicher Str.~77, 50937 K\"oln, Germany}
}
\maketitle

% Abstract

\begin{abstract}
We present a novel extrapolation scheme for high order series
expansions. The idea is to express the  series, obtained in orders
of an external variable, in terms of an internal parameter of the
system. Here we apply this method to the 1-triplet dispersion in
an antiferromagnetic $S=\frac{1}{2}$ Heisenberg ladder. By the use
of the internal parameter the accuracy of the truncated series  is
enhanced tremendously.
\end{abstract}

%%% Nimm PACS Nummern wie vorher in unseren frueheren Arbeiten, eventuel
%%% etwas Materialbezogenes raus lassen und dafuer etwas wie "Methoden der
%%% theoretischen Physik" rein nehmen.

% The main text

\section{Introduction}
High order series expansions have become a powerful tool in the field of
strongly correlated electron systems \cite{cit0}.
Especially in low dimensional quantum
spin systems significant progress has been achieved,
 see e.g.\ [2-6].
Usually the obtained truncated series
 must be extrapolated using various extrapolation schemes in order to obtain
 results for the physically interesting regions \cite{cit6}.
Albeit these extrapolations are a standard technique they always
introduce some uncertainty about the results. Therefore, a general
transformation scheme which allows to read off the information
content of the high order series more directly is highly
desirable. Here we will propose such a general scheme. Its
usefulness will be demonstrated for the 1-triplet dispersion in
antiferromagnetic 2-leg $S=\frac{1}{2}$ Heisenberg ladders. The
series was
 obtained previously by various techniques \cite{cit3,cit4,cit2};
 we used a perturbative continuous unitary transformation (CUT)
\cite{cit1}.

The Hamiltonian for the antiferromagnetic 2-leg Heisenberg ladder reads
\begin{equation}
  \label{H_start}
   H = \sum_{i}\left[ J_{\parallel}\left( {\bf S_{1,i}S_{1,i+1}}+
{\bf S_{2,i}S_{2,i+1}}\right) + J_{\perp}{\bf S_{1,i}S_{2,i}} \right] \,
\end{equation}
where $J_{\parallel}$ and $J_{\perp}$ are the exchange couplings
on the legs and on the rungs,
respectively. The
 subscript $i$ denotes the rungs and the subscript $1,2$ the legs of
the ladder.\\
We use a CUT to map the Hamiltonian $H$ to an effective Hamiltonian
 $H_{\rm eff}$ which conserves the number of rung-triplets, i.e.\
$[H_{\rm 0},H_{\rm eff}]=0$ where $H_{\rm 0}:=H|_{J_{\rm
\parallel}=0}$ \cite{cit1}. The ground state of $H_{\rm eff}$ is
the rung-triplet vacuum. The effective Hamiltonian $H^{\rm eff}$
is calculated in the 1-triplet subspace of the Hilbert space to
order $14$ in $x:=J_{\parallel}/J_{\perp}$. The ground state
energy $E_0=\left< 0|H_{\rm eff}|0\right>$ and the 1-triplet
dispersion $\omega(k)=\left< k|H_{\rm eff}|k\right> - E_0$ is
obtained \cite{cit3,cit4,cit2}. The 1-triplet dispersion is
expressed in terms of the 1-triplet hopping
 amplitudes $t_n(x)=\left\langle i|H_{\rm eff}|i+n\right\rangle$ where
$|i\rangle$ denotes the state
 with one triplet on rung $i$
\begin{equation}
 \label{Disp}
 \omega(k)/J_{\perp} = \sum_n t_n(x)\cos(nk) \quad .
\end{equation}
\begin{figure}[b]
\label{GapFig}
\begin{center}
\includegraphics[width=5cm,height=5cm]{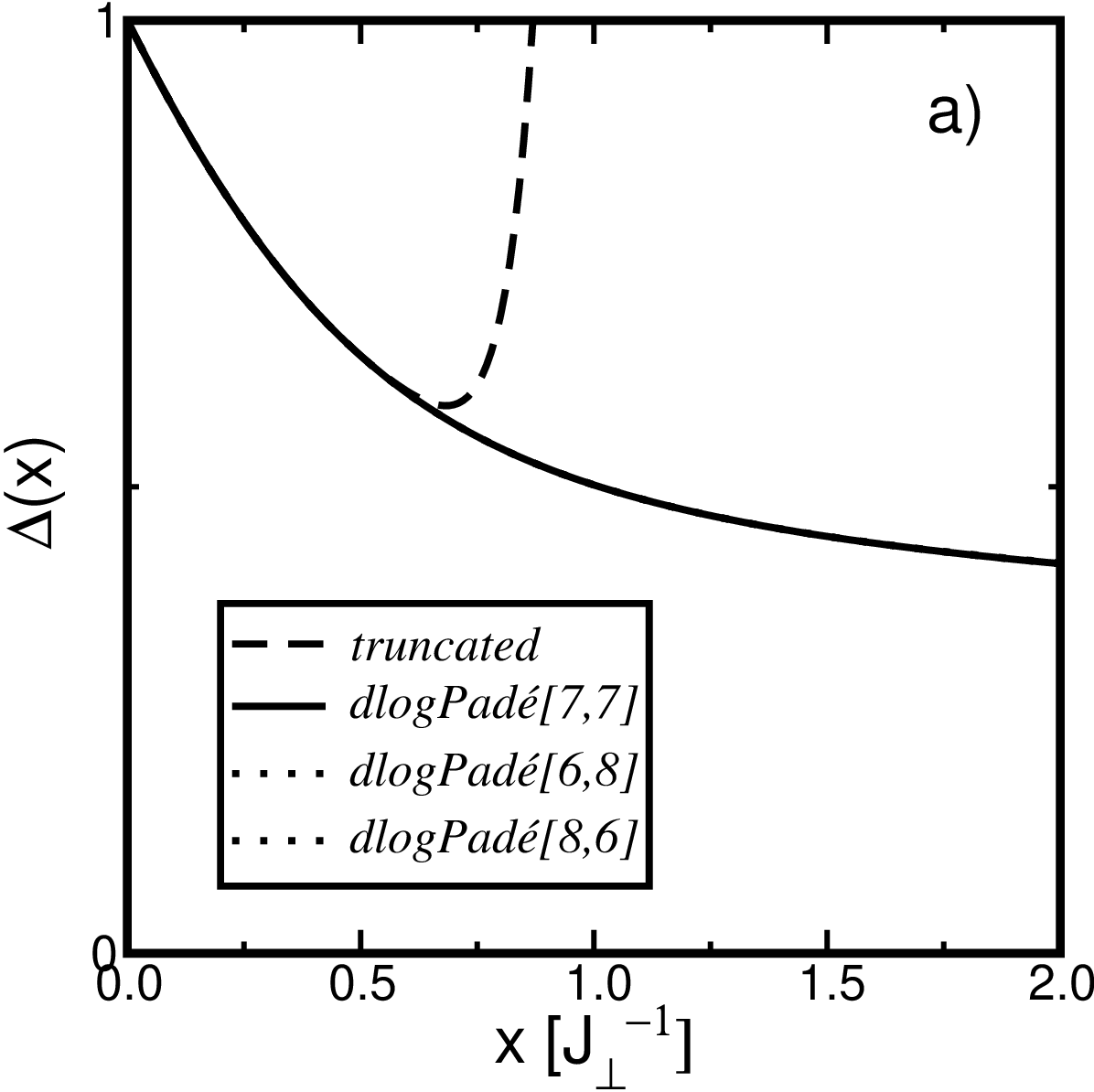}
\includegraphics[width=5cm,height=5cm]{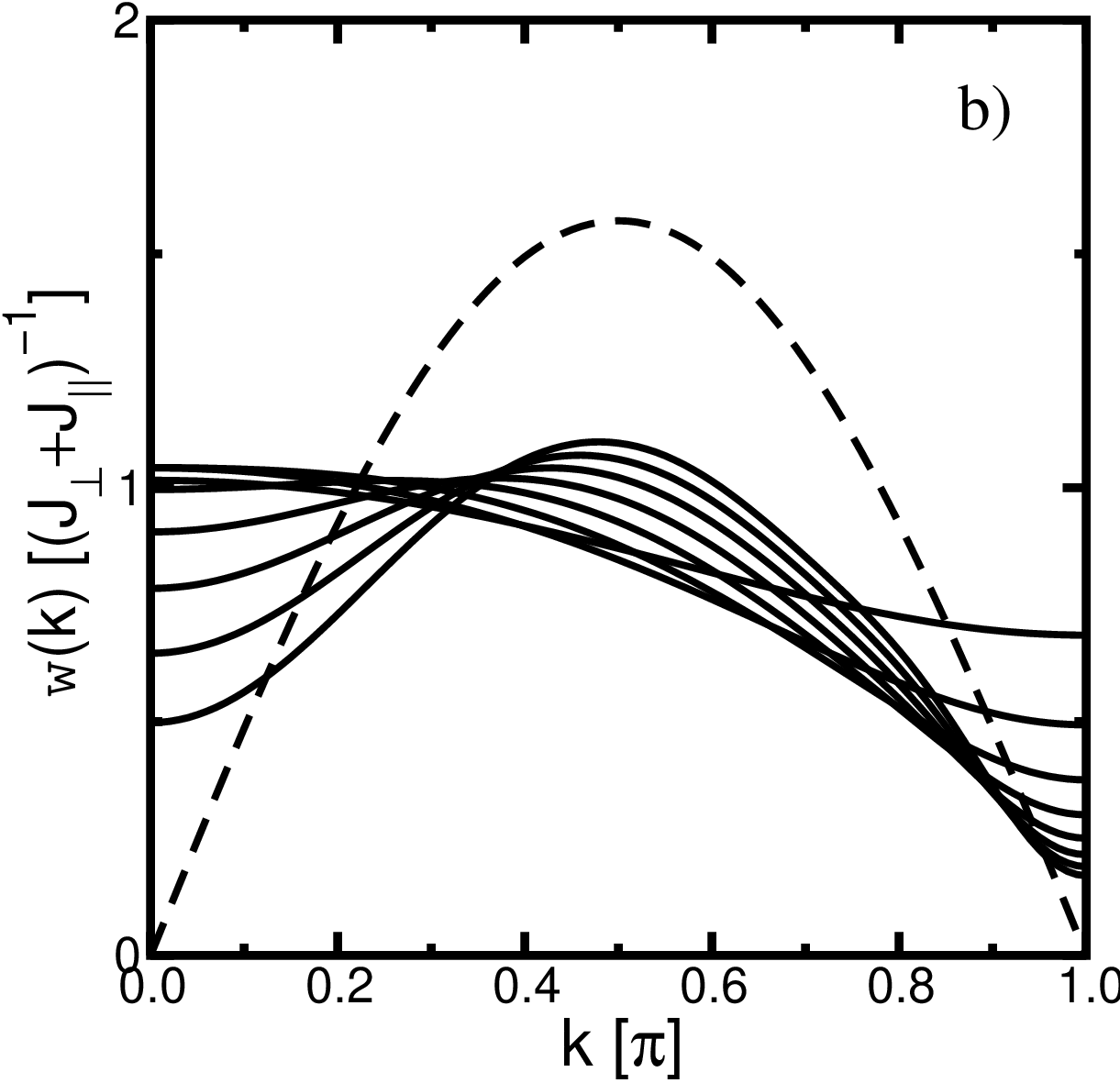}
\end{center}
\caption{(a) Gap $\Delta(x)/J_\perp$. The solid and the
dotted curves curves result from a [7,7], [6,8] and a [8,6] dlogPad\'e
approximant, resp., assuming constant behavior on $x\to\infty$.
The long-dashed curve depicts the truncated series.\newline
%%% Noch an die tatsaechliche Abbildung anpassen und Papier-Skript
%%% beruecksichtigen.
(b) Dispersion relation $\omega(k)$ in units of
$J_\perp+J_\parallel$ for the couplings
$x=J_\parallel/J_\perp=0.2, 0.4,\ldots, 1.6$. The highest plot at
$k=\pi$ is $x=0.2$ and the lowest plot is $x=1.6$. The long-dashed
curve is the limit of isolated chains ($x=\infty$), i.e.\ the des
Cloizeaux-Pearson result $(\pi/2)\sin(k)$ \cite{cit7}.
}
\end{figure}
The 1-triplet dispersion has a minimum for $k=\pi$, i.e.\
 the 1-triplet gap $\Delta(x)$ given by $\omega(\pi)$.
In Fig.~1a, the results for $\Delta(x)$ are shown. The truncated
 series for $\Delta(x)$ yields a satisfactory agreement up to
$x\approx 0.6$. Beyond this value it cannot be used as estimate
for the value of the gap. More sophisticated extrapolation
schemes, however, still work fine.\newpage
\section{Extrapolation Scheme and Results}
Generically, the various physical quantities in a given system
depend in a complicated way on the external control parameters.
Expanding the quantities under study in terms of one of the
external control parameters, let us say $x$, yields the bare,
truncated series which can only rarely be directly used
to compute the quantities. This is so since singularities
induced by phase transitions easily spoil the convergence
of the series. For instance, a correlation length
diverges and the corresponding energy gap closes rendering
an expansion about the gapped phase difficult.

If the convergence of the series is deteriorated due to an
incipient phase transition it is reasonable to assume that {\it
all} quantities in the particular system behave in a similar
fashion. If this is so we may proceed in two separate steps: (i)
We extrapolate an internal parameter which may serve as a measure
of the distance to the phase transition as reliably as we can.
Thereby, we can attribute reliably to a given value of $x$ the
corresponding value of the internal parameter. (ii) We express all
{\it other} quantities as functions of the internal parameter.
According to our argument the latter functional dependencies are
expected to be much simpler, i.e.\ they are much less singular.
The canonical candidate for the internal parameter measuring the
distance to a phase transition or, more generally, to some
singular situation is the energy gap $\Delta$. It is inversely
proportional to the correlation length $\xi$ which plays the role
of the internal control parameter in standard renormalization
group treatments.

In the following, we illustrate for the 1-triplet dispersion of
the model in (\ref{H_start}) that this scheme works indeed
stunningly well. The gap $\Delta(x)$ can be extrapolated reliably
up to $x\approx 2$ using dlogPad\'e-approximants, Fig.~1a. In this
extrapolation we can exploit additional properties of the gap such
as its positivity and its asymptotic behavior for $x\to\infty$. In
this way, very reliable extrapolations are possible (Fig.~1a) so
that step (i) is successfully solved. Note that $\Delta(x)$ in
units of $J_\perp$ does not vanish on $x\to\infty$ but rests
finite \cite{shelt96,greve96}.

For step (ii) we define
\begin{equation}
 \label{G}
 p(x)=1-\Delta(x)/((1+x)J_\perp) =  1- \Delta(x)/(J_\parallel+J_\perp) \
 .
\end{equation}
 In  units of $J_\parallel+J_\perp$ the gap is unity at $x=0$ and it
goes to zero on $x\to\infty$. So $p(x)$ varies monotonically
between 0 and 1 when $x$ is increased from 0 to $\infty$. Note
that $p=1$, i.e.\ $x=\infty$, constitutes the limit where the spin
ladder becomes a system of two isolated spin chains. Since one has
$p\propto x$ for small $x$ any expansion in $x$ can be rewritten
as expansion in $p$ of the same order as the series in $x$, yet
with other coefficients! This is done by inverting Eq.~\ref{G},
thus completing the second step.

We applied the extrapolation scheme proposed above to the
expansions of the 1-triplet hopping amplitudes $t_n$ in
Eq.~\ref{Disp} which yields the 1-triplet dispersion
$\omega(k,p)$. In Fig.~1b the results are depicted for the
truncated series in $p$ without any further extrapolation! For $x<
0.6$
%%% (Es muss einen Befehl der Art "lessapprox" geben,
%%% suche bitte mal)
the dispersion is a monotonic decreasing
 function in $k$ whereas for larger values of $x$ a characteristic dip  at
$k=0$ appears. For comparison, the limiting case of isolated
chains is also included \cite{cit7}. The main point is that the
truncated series in $p$ gives a quantitatively correct 1-triplet
dispersion up to $x\approx 1.2$ (cf.\ Ref.\ \cite{cit3}) and
qualitatively good results up to $x\approx 2$ while the truncated
series in $x$ can be trusted only up to $x\approx 0.6$.

\section{Conclusion}
We presented a novel generally applicable extrapolation scheme for
high order series expansions. The basic idea is to use an {\it
internal} control parameter instead of an {\it external} control
parameter as expansion variable. All difficulties stemming from
singularities are dealt with in the determination of the
dependence of the internal parameter on the external one.

The power of the  novel scheme was demonstrated  for the 1-triplet
dispersion of the antiferromagnetic $S=\frac{1}{2}$
 2-leg Heisenberg ladder. Further investigations on other quantities and systems are in
progress. In addition, the use of standard extrapolation
techniques on the series in the internal parameter constitutes a
very promising route to extend the applicability of strong
coupling expansions.

We are indebted to H.~Monien and E.~M\"uller-Hartmann for helpful
discussions and to the DFG for financial support in SP 1073 and in
SFB 608.

\end{document}